\documentclass[aps,prc,superscriptaddress,twocolumn,showpacs,nofootinbib]{revtex4-1}
\usepackage[utf8]{inputenc}
\usepackage{graphicx}   
\usepackage{latexsym}   
\usepackage{enumerate}
\usepackage{rotating,booktabs,multirow}
\usepackage{amsmath}
\usepackage{amsfonts}
\usepackage{amssymb}
\usepackage{multirow}
\usepackage{bm}
\usepackage{xcolor}
\usepackage[colorlinks=true,linktocpage=true,linkcolor=blue,citecolor=blue,allcolors=blue]{hyperref}
\usepackage{url}

\usepackage{float}

\begin{document}
\title{Directed and elliptic flow of light nuclei and hypernuclei in Au+Au collisions at $\sqrt{s_\mathrm{NN}}=3$ GeV: Coalescence vs. Statistical Fragmentation}

\author{Tom Reichert}
\affiliation{Institut f\"ur Theoretische Physik, Goethe-Universit\"at Frankfurt, Max-von-Laue-Strasse 1, D-60438 Frankfurt am Main, Germany}
\affiliation{Frankfurt Institute for Advanced Studies (FIAS), Ruth-Moufang-Str.1, D-60438 Frankfurt am Main, Germany}
\affiliation{Helmholtz Research Academy Hesse for FAIR (HFHF), GSI Helmholtz Center for Heavy Ion Physics, Campus Frankfurt, Max-von-Laue-Str. 12, 60438 Frankfurt, Germany}

\author{Manjunath Omana Kuttan}
\affiliation{Frankfurt Institute for Advanced Studies (FIAS), Ruth-Moufang-Str.1, D-60438 Frankfurt am Main, Germany}

\author{Apiwit Kittiratpattana}
\affiliation{Suranaree University of Technology, University Avenue 111, Nakhon Ratchasima 30000, Thailand}

\author{Nihal~Buyukcizmeci}
\affiliation{Department of Physics, Selcuk University, 42079 Kamp\"us, Konya, T\"urkiye}

\author{Alexander Botvina}
\affiliation{Institut f\"ur Theoretische Physik, Goethe-Universit\"at Frankfurt, Max-von-Laue-Strasse 1, D-60438 Frankfurt am Main, Germany}
\affiliation{Helmholtz Research Academy Hesse for FAIR (HFHF), GSI Helmholtz Center for Heavy Ion Physics, Campus Frankfurt, Max-von-Laue-Str. 12, 60438 Frankfurt, Germany}

\author{Jan Steinheimer}
\affiliation{GSI Helmholtzzentrum f\"ur Schwerionenforschung GmbH, Planckstr. 1, 64291 Darmstadt, Germany}
\affiliation{Frankfurt Institute for Advanced Studies (FIAS), Ruth-Moufang-Str.1, D-60438 Frankfurt am Main, Germany}

\author{Marcus Bleicher}
\affiliation{Institut f\"ur Theoretische Physik, Goethe-Universit\"at Frankfurt, Max-von-Laue-Strasse 1, D-60438 Frankfurt am Main, Germany}
\affiliation{Helmholtz Research Academy Hesse for FAIR (HFHF), GSI Helmholtz Center for Heavy Ion Physics, Campus Frankfurt, Max-von-Laue-Str. 12, 60438 Frankfurt, Germany}

\begin{abstract}
The harmonic flow coefficients of light nuclei and hypernuclei in Au+Au collisions at $\sqrt{s_\mathrm{NN}}=3$ GeV are investigated using the Ultra-relativistic Quantum Molecular Dynamics transport model. For the Equation-of-State we employ a density and momentum dependent potential from the Chiral-Mean-Field model. Light nuclei and hypernuclei production is described at kinetic freeze-out via a coalescence mechanism or with a statistical multi-fragmentation calculation. The directed flow $v_1$ of p, d, t, $^3$He, $^4$He as well as the $\Lambda$, $^3_\Lambda$H and $^4_\Lambda$H is shown to approximately scale with mass number $A$ of the light cluster in both calculations. This is in agreement with the experimental results for the directed flow measured by STAR. Predictions for the directed and elliptic flow of (hyper)nuclei at further RHIC-FXT and FAIR energies show that the scaling properties should improve as the beam energy is increased.
\end{abstract}

\maketitle

\section{Introduction}
Quantum-Chromo-Dynamics (QCD), the theory of the strong interaction, is typically studied in hadron+hadron or nucleus+nucleus collisions. In addition these laboratory based investigations can be supplemented by measurements of astrophysical objects such as compact stellar objects, e.g. neutron stars or binary neutron star mergers. While the two systems exist on vastly different time and length scales their connections are nevertheless manifold. The baryon densities reached in heavy-ion collisions coincide with the densities present inside neutron stars. This provides a bridge in terms of the nuclear Equation-of-State (EoS) \cite{Most:2022wgo}. The observation of neutron stars with large masses $M>2M_\odot$ and the slow variation of neutron star radii above $M>1.4M_\odot$ suggests that the Equation-of-State is rather stiff at large baryon densities favoring large values of the speed-of-sound $c_s^2$. The apparent softening due to the presence of hyperons, $Y$, in the neutron star core \cite{Ozel:2010bz,Bonanno:2011ch,Lastowiecki:2011hh,Blaschke:2015uva} is contradictory to these findings. The $YN$ and three body $YNN$ interactions (where $N$ is a nucleon) which could lead to a stiffening of the EoS are thus of utmost importance for the understanding of neutron stars with masses $M>2M_\odot$. Theoretical chiral effective field theory (EFT) calculations are able to provide estimates for the $YN$ interaction strength \cite{Haidenbauer:2013oca,Haidenbauer:2019boi,Haidenbauer:2021wld}, while estimates of the binding energies can be calculated from effective theories \cite{Friedman:2022bpw,Jinno:2023xjr}. An excellent opportunity for the investigation of the hyperon-nucleon interaction is provided by hypernuclei, i.e. bound objects consisting of hyperons and nucleons.

Traditionally hypernuclei have been studied in bubble chamber experiments \cite{doi:10.1080/14786440308520318}. Nowadays hypernuclei and thus the hyperon-nucleon interaction can be studied in heavy-ion collision experiments at relativistic energies which provide a sufficient source of both strangeness and light clusters. Hypernuclei production has been measured at RHIC \cite{STAR:2017gxa,STAR:2019wjm,STAR:2021orx} and LHC energies \cite{ALICE:2015oer,ALICE:2019vlx} and recently observed at GSI's HADES experiment \cite{HADES:2022gms,Spies:2023cxi}. Thermal model estimates suggest that the highest hypernuclei multiplicities can be expected around $\sqrt{s_\mathrm{NN}} \approx 4-6$ GeV \cite{Andronic:2010qu}, which is the energy range that is going to be probed with the high-luminosity FAIR facility, allowing for very precise measurements of hypernuclei properties (harmonic flow, correlations, ...). The first benchmarking measurement of harmonic flow of hypernuclei has been recently achieved by the STAR collaboration \cite{STAR:2022fnj} at BNL's RHIC collider. STAR's measurement of the directed flow of the hypertriton ($^3_\Lambda$H) and the hyperhydrogen-4 ($^4_\Lambda$H) might allow to further constrain the hyperon-nucleon interaction and opens the route for further measurements and theoretical investigations at STAR-FXT and the upcoming FAIR facility.

On the theoretical side the description of hypernuclei formation and their interaction with the bulk medium in heavy-ion collisions requires a dynamical microscopic transport model. Based on measured cross sections transport models provide well constrained physical simulations of $AA$ collisions. The recent measurement of the $\Lambda p$ elastic scattering cross section \cite{CLAS:2021gur} complement the rather scarce data on $\sigma_{\Lambda p}$ and allow to further refine the treatment of hyperons in transport models. Hypernuclei production has been studied in several transport approaches \cite{Glassel:2021rod,Reichert:2022mek}, thermal model calculations \cite{Andronic:2010qu} and in 3-fluid models \cite{Kozhevnikova:2024itb}.

In this work the directed and elliptic flow of the protons p, the light clusters d, t, $^3$He and $^4$He and of the $\Lambda$ and the hypernuclei $^3_\Lambda$H and $^4_\Lambda$H is studied with the UrQMD model. We supplement UrQMD with two approaches for the calculation of (hyper)clusters:

\begin{enumerate}
\item The phase-space coalescence mechanism \cite{Gyulassy:1982pe,Mattiello:1996gq,Sombun:2018yqh} 
\item A statistical multi-fragmentation (SMM) approach \cite{Bondorf:1995ua}.
\end{enumerate}

\section{Model setup}
\subsection{UrQMD}
The Ultra-relativistic Quantum Molecular Dynamics model (UrQMD, v3.4) \cite{Bass:1998ca,Bleicher:1999xi,Bleicher:2022kcu} is a microscopic transport model based on the covariant propagation of hadrons as explicit degrees of freedom. As a QMD type model UrQMD explicitly propagates all n-body correlations on an event-by-event basis, an important ingredient to calculate the production of clusters. Hadron+hadron interactions are implemented via the geometric interpretation of their cross section which is taken from empirical data, if available, or derived from effective models. In the investigated energy regime the binary elastic and inelastic reactions are supplemented by potential interactions for which the employed density and momentum dependent potential is derived from the Chiral-Mean-Field (CMF) model \cite{Steinheimer:2024eha}. This parametrization of the CMF EoS was shown to be consistent with astrophysical observations and is able to describe measured flow data from the HADES experiment \cite{Steinheimer:2025hsr}. 

For previous studies on cluster and hypernuclei production using UrQMD, we refer the reader to \cite{Gaebel:2020wid,Kittiratpattana:2022knq,Buyukcizmeci:2020asf,Reichert:2022mek,Buyukcizmeci:2023azb}, and for studies of flow coefficients in the present energy regime we refer to \cite{Reichert:2024ayg}.

\subsection{Phase-space coalescence}
In UrQMD cluster production is modeled via the coalescence approach, a well established mechanism for the production of clusters. In the original momentum-space coalescence approach \cite{Butler:1963pp} and in the generalized phase space coalescence approach used here, the invariant spectra of light clusters and hypernuclei are related to the invariant spectra of their respective constituents. Due to the large average momentum transfer of binary interactions in the medium in comparison to the binding energies of the cluster, the formation of bound objects occurs at the kinetic freeze-out hyper surface. The coalescence approach is justified because cluster formation within the hadronic medium is strongly suppressed due to disintegration of the clusters by any interaction. Coalescence in UrQMD is implemented as described in detail in Refs. \cite{Sombun:2018yqh,Hillmann:2018nmd,Hillmann:2019wlt,Hillmann:2021zgj,Reichert:2022mek}. It should be noted that the UrQMD phase space coalescence approach has been recently compared to deuteron formation with potential interactions in \cite{Kireyeu:2022qmv} yielding very similar results. 

As the inclusion of QMD interactions leads to strongly correlated clusters (see e.g. \cite{Kireyeu:2023spj}) at the freeze-out, the coalescence parameters $\Delta r_\mathrm{max}$ and $\Delta p_\mathrm{max}$ need to be refitted. For the light nuclei and hypertriton this was done to describe the measured STAR midrapidity yields, while for the hyperhydrogen-4, the same parameters as for the hypertriton are used. The parameters for this study are summarized in Table \ref{tab:fac}

\begin{table}[h]
\centering
\begin{tabular}{|c|c|c|c|c|c|}
\hline
 & deuteron  & $^{3}$H or $^{3}$He & $^{4}$He & $_{\Lambda}^{3}$H &  $_{\Lambda}^{4}$H \\
\hline
spin-isospin & 3/8 & 1/12 & 1/96 & 1/12 & 1/96\\
\hline
$\Delta r_\mathrm{max}$ [fm] & 4.0 & 3.5 & 3.5  & 9.5 & 9.5 \\
\hline
$\Delta p_\mathrm{max}$  [GeV] & 0.33 & 0.45 & 0.55 & 0.15 & 0.25 \\
\hline
\end{tabular}
\caption{Probabilities and parameters used in the UrQMD  phase-space coalescence for the QMD mode.\label{tab:fac}}
\end{table}

\subsection{Statistical multi-fragmentation} 
Another way to take into account the interaction of baryons leading to the formation of nuclei is to use the statistical approach. Similar to the coalescence model we consider the baryons which are close in phase space. However, as established in numerous studies of the multifragmentation reaction the baryons can form nuclei if they are in the coexistence region of the nuclear liquid-gas type phase transition (see, e.g., \cite{Bondorf:1995ua,Buyukcizmeci:2023azb,Botvina:2020yfw} and references therein). This region is characterized by local chemical equilibrium, it has relative low temperature (around $T \approx 5-10$ MeV) and the corresponding density is around 0.1--0.3 $\rho_0$ ($\rho_0 \approx 0.15$ fm$^{-3}$ being the ground state nuclear density). In this case statistical models can be used for description of the fragmentation in nuclear matter \cite{Bondorf:1995ua}. Therefore, with the coalescence-like prescription \cite{Botvina:2014lga} we select baryons (i.e., subdivide the expanding matter) into several local primary nuclear clusters which are characterized by the above mentioned parameters. The excitation energy of such clusters is around 10--15 MeV per nucleon. In the following we describe the nucleation process inside these primary clusters as their statistical decay within the statistical multifragmentation model (SMM) \cite{Bondorf:1995ua}. The parameters of the model were taken the same as in Ref. \cite{Buyukcizmeci:2023azb}. The model has been used for a variety of systems and collision energies providing qualitatively good results \cite{Buyukcizmeci:2020asf,Botvina:2020yfw}. The extension to the strange sector, respectively to hypernuclei, is described in detail in Ref. \cite{Botvina:2007pd}. The combined UrQMD+SMM framework has recently demonstrated its capabilities to describe the light nuclei and hypernuclei multiplicities and spectra in Au+Au collisions at $\sqrt{s_\mathrm{NN}}=3$ GeV \cite{Buyukcizmeci:2023azb} and is therefore well suited for the present analysis of harmonic flow.

\begin{figure}[t]
    \centering
    \includegraphics[width=\columnwidth]{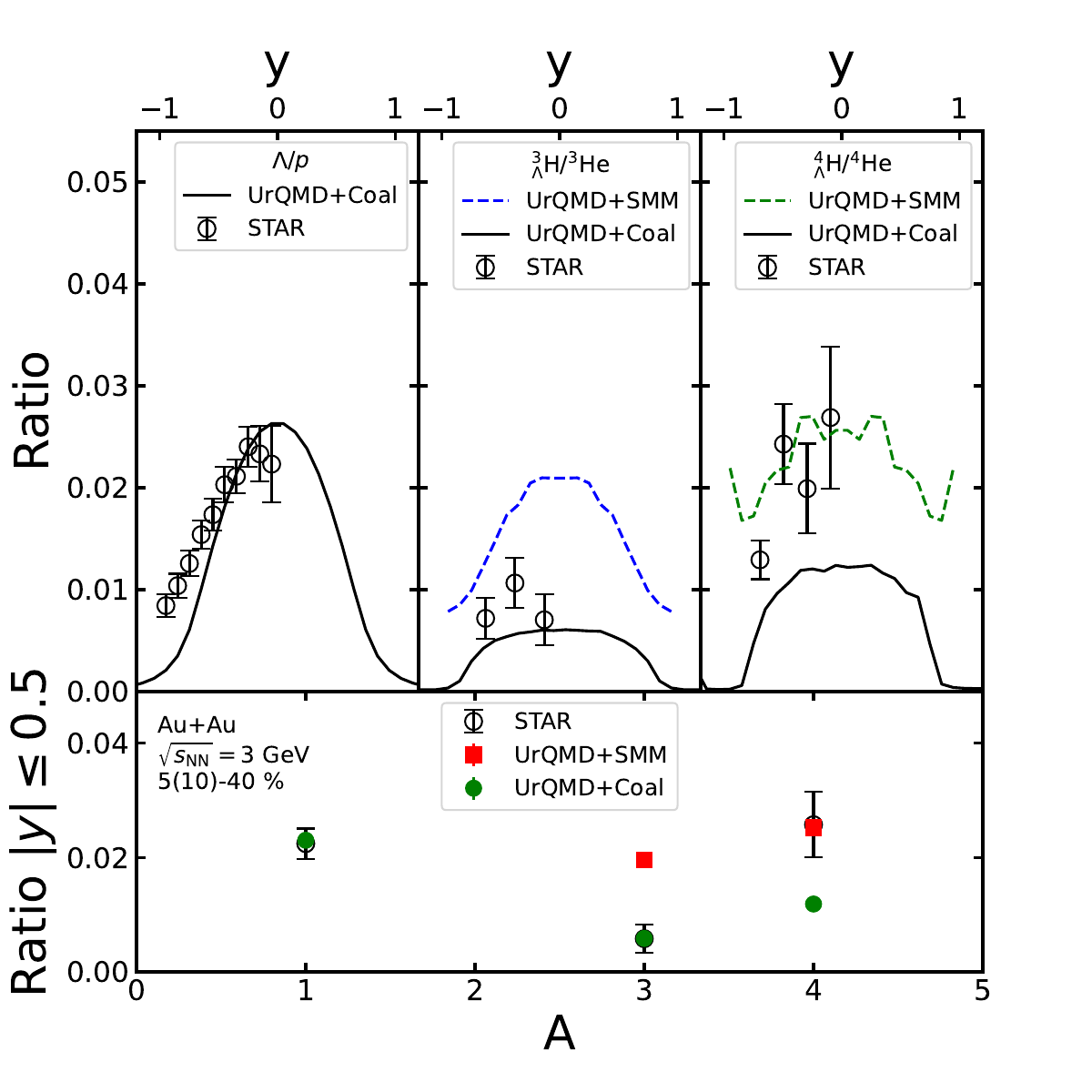}
    \caption{[Color online] The rapidity dependence of the ratio of $\Lambda / p$ (left), $^3_\Lambda$H/$^3$He (middle) and $^4_\Lambda$H/$^4$He (right) from UrQMD with coalescence (solid lines) and from UrQMD with subsequent statistical multi-fragmentation (dashed lines for the hypernuclei). The results are compared to data from STAR \cite{STAR:2021orx,STAR:2024znc}. The lower panel shows the ratio integrated over the midrapidity interval $|y|\leq0.5$ as a function of mass number $A$.}
    \label{fig:ratio}
\end{figure}

\section{Results}
For the comparison with the STAR data we simulated 10-40\% central (in case of the light nuclei) and 5-40\% central (in case of the hypernuclei) events of Au+Au collisions at $\sqrt{s_{\rm NN}}=3$~GeV center of mass energy with the UrQMD model in line with the centralities used by the STAR experiment. The centrality is selected via impact parameter cuts following a Glauber model as it is done for the experiments. For the calculation of light (hyper-)nuclei flow, the light (hyper-)nuclei are either produced with coalescence model at the kinetic freeze-out, i.e. after the last interaction of each cluster constituent or using the statistical multi-fragmentation model. For the formation of light (hyper-)fragments within the SMM the UrQMD simulations are stopped at a fixed time (here: t=40 fm/c). The coalescence parameter for the primary hot cluster recognition in the SMM model has been set to $v_c=0.22$. Both the $v_c$ parameter and the fixed time input have previously shown to provide good results \cite{Botvina:2014lga,Buyukcizmeci:2020asf,Botvina:2020yfw} for the multiplicities and spectra of light nuclei and hypernuclei.

\subsection{Multiplicities and ratios of the hypernuclei}
We start our investigation by calculating the rapidity dependent ratios of $\Lambda / p$, $^3_\Lambda$H/$^3$He and $^4_\Lambda$H/$^4$He. Fig. \ref{fig:ratio} shows the rapidity dependent ratios of $\Lambda  / p$ (left), $^3_\Lambda$H/$^3$He (middle) and $^4_\Lambda$H/$^4$He (right) from UrQMD with coalescence (solid lines) and from UrQMD with subsequent statistical multi-fragmentation (SMM, dashed lines for the hypernuclei). The lower panel shows the ratio integrated over the midrapidity interval $|y|\leq0.5$ as a function of mass number $A$. The results are compared to data from STAR (open symbols) \cite{STAR:2021orx,STAR:2024znc}. 

It can be seen that UrQMD provides a good description of the hyperon yield ratios, which is an important input for the subsequent cluster formation. The coalescence and multifragmentation processes provide both a satisfactory description of the hypernuclei production, while the yield ratios in the SMM model are generally higher\footnote{Note that the $\Lambda/p$ and $^4_\Lambda$H/$^4$He ratios are similar in magnitude, while the $^3_\Lambda$H/$^3$He is suppressed by a factor of approx. 2. This might be attributed to excited $^4_\Lambda$H$^*$ states.}. With the ratios of rapidity spectra being reasonable well reproduced we can now move on to the first two flow coefficients.

\subsection{Flow extraction}
The harmonic flow coefficients $v_n$ arise from the Fourier series expansion of the azimuthal angular distribution $\mathrm{d}N/\mathrm{d}\phi$ with respect to the reaction plane $\Psi_\mathrm{RP}$
\begin{equation}
    \frac{\mathrm{d}N(y,p_\mathrm{T})}{\mathrm{d}\phi} \propto 1 + 2\sum\limits_{n=1}^\infty v_n(y,p_\mathrm{T}) \cos(n(\phi-\Psi_\mathrm{RP})).
\end{equation}
Experimentally the angle of the reaction plane $\Psi_\mathrm{RP}$ is unknown and has to be estimated from the correlations in the data. At much higher energies the flow coefficients are therefore extracted with respect to the n-th order event planes or are calculated by means of two- or four-particle cumulants. The STAR collaboration, however, used the first order event plane to reconstruct the directed flow in their recent analysis of hypernuclei flow \cite{STAR:2022fnj} which provides a good approximation for the reaction plane in the respective energy regime. In the following results, the flow coefficients are therefore extracted with respect to the reaction plane which is known in the simulation. Thus, $\Psi_\mathrm{RP}=0$ and $v_n = \langle \cos(n\phi) \rangle$, in which the average $\langle\cdot\rangle$ is the inclusive ensemble average and $\tan(\phi) = p_y/p_x$.

\begin{figure}[t]
    \centering
    \includegraphics[width=\columnwidth]{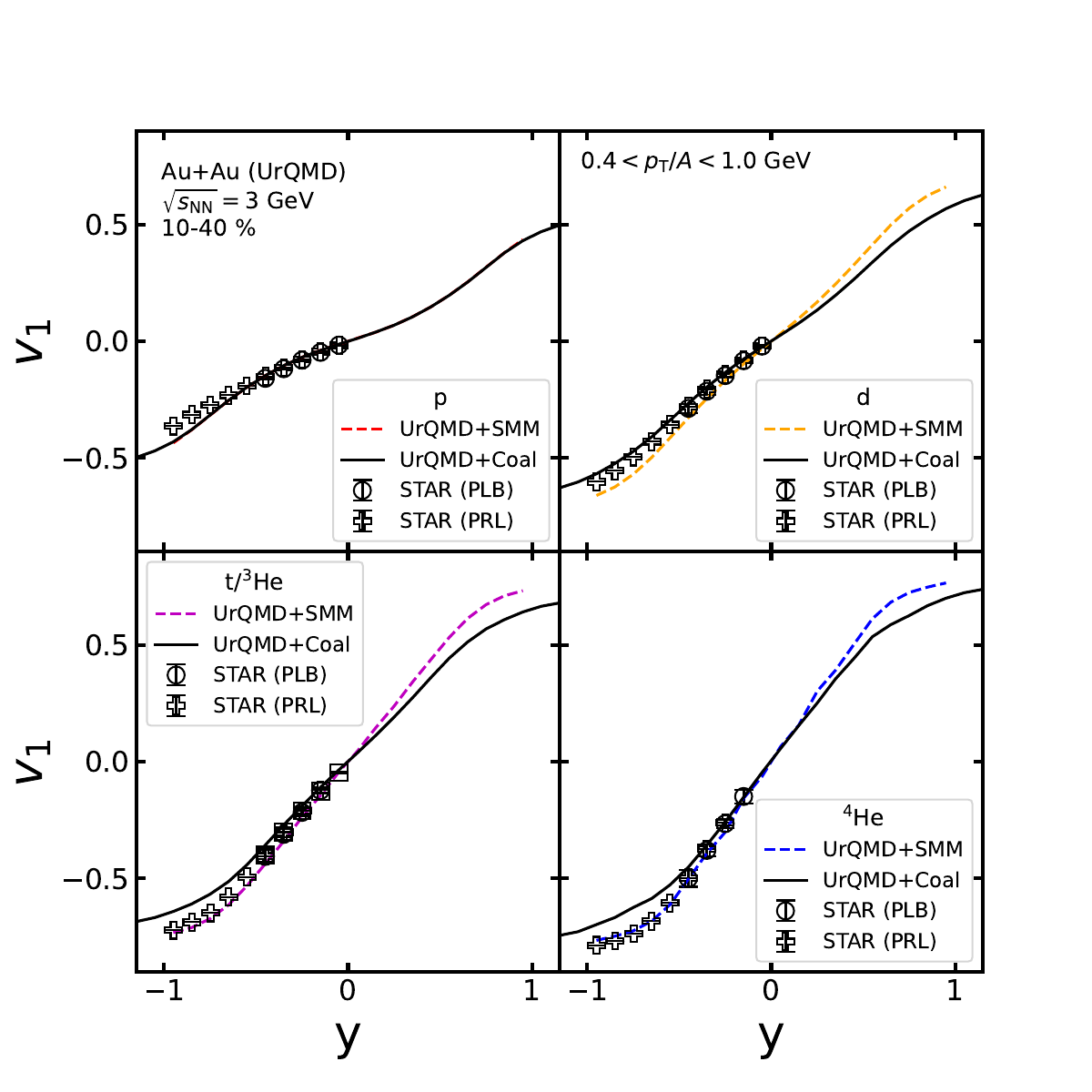}
    \caption{[Color online] The directed flow $v_1$ as a function of rapidity of protons (upper left), deuterons (upper right), tritons and $^3$He (lower left) and $^4$He (lower right) from 10-40\% central Au+Au collisions at $\sqrt{s_\mathrm{NN}}=3$ GeV from UrQMD with coalescence (solid lines) and from UrQMD combined with the statistical multi-fragmentation model (dashed lines). Experimental data points are taken from STAR \cite{STAR:2021ozh,STAR:2021yiu,STAR:2022fnj}.}
    \label{fig:v1_y_light}
\end{figure}
\begin{figure}[t]
    \centering
    \includegraphics[width=\columnwidth]{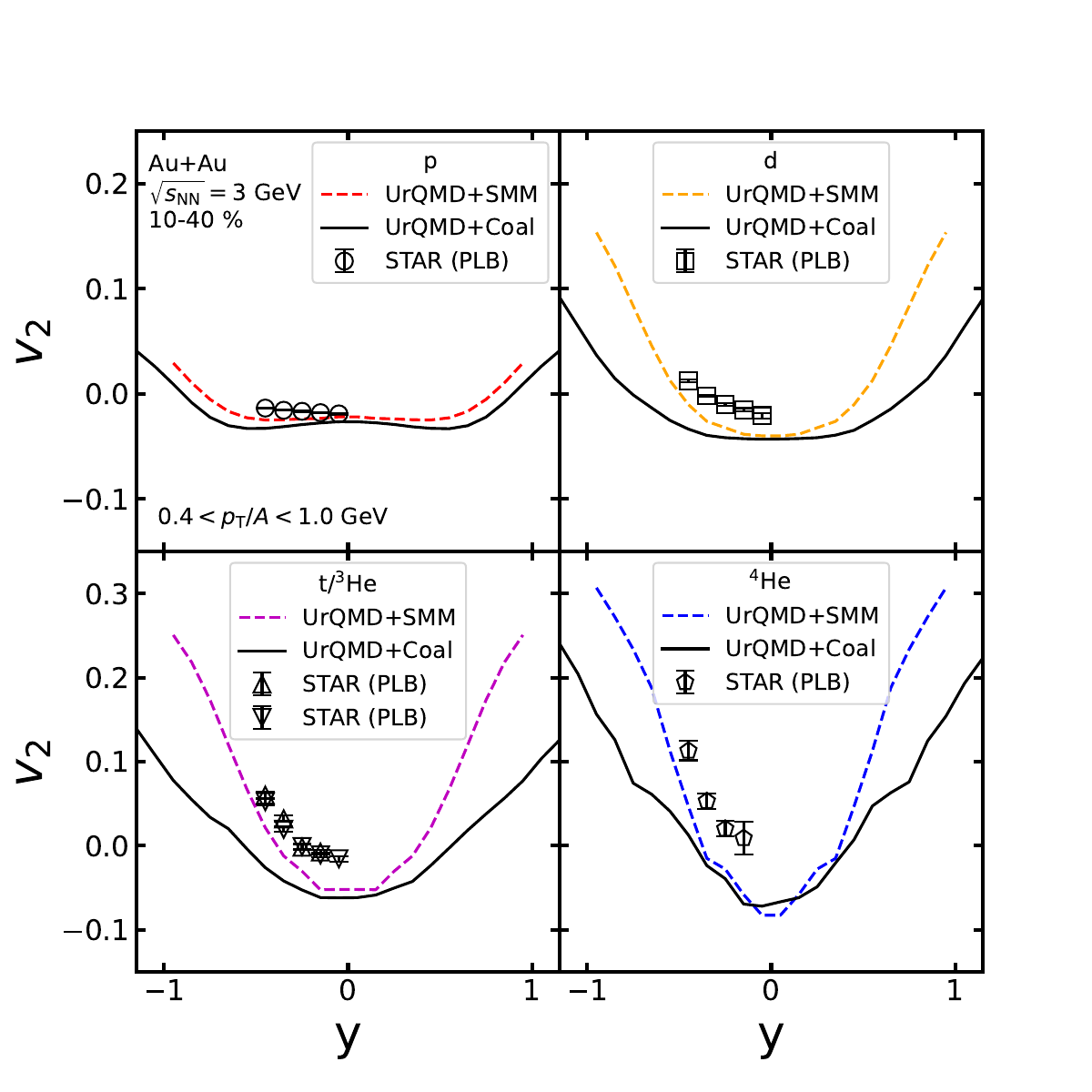}
    \caption{[Color online] The elliptic flow $v_1$ as a function of rapidity of protons (upper left), deuterons (upper right), tritons and $^3$He (lower left) and $^4$He (lower right) from 10-40\% central Au+Au collisions at $\sqrt{s_\mathrm{NN}}=3$ GeV from UrQMD with coalescence (solid lines) and from UrQMD combined with the statistical multi-fragmentation model (dashed lines). Experimental data points are taken from STAR \cite{STAR:2021ozh,STAR:2021yiu}.}
    \label{fig:v2_y_light}
\end{figure}

\begin{figure}[t]
    \centering
    \includegraphics[width=\columnwidth]{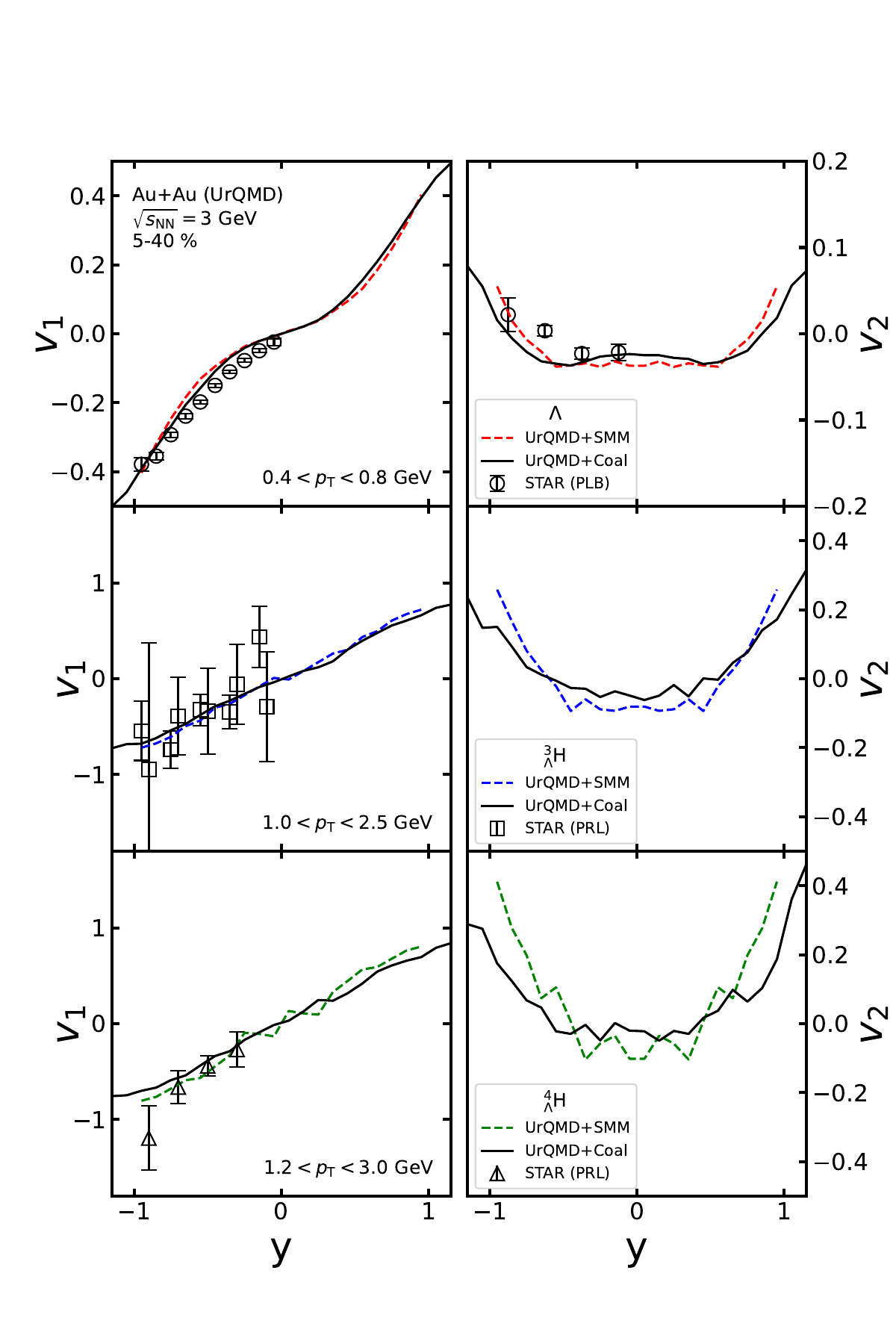}
    \caption{[Color online] The directed flow $v_1$ (left column) and the elliptic flow $v_2$ (right column) as a function of rapidity of the $\Lambda$ (upper row), $^3_\Lambda$H (middle row) and $^4_\Lambda$H (lower row) from 5-40\% central Au+Au collisions at $\sqrt{s_\mathrm{NN}}=3$ GeV from UrQMD with coalescence (solid lines) and from UrQMD combined with the statistical multi-fragmentation model (dashed lines). Experimental data points are taken from \cite{STAR:2021yiu,STAR:2022fnj}.}
    \label{fig:v1_v2_y_hyper}
\end{figure}

\begin{figure}[t]
    \centering
    \includegraphics[width=\columnwidth]{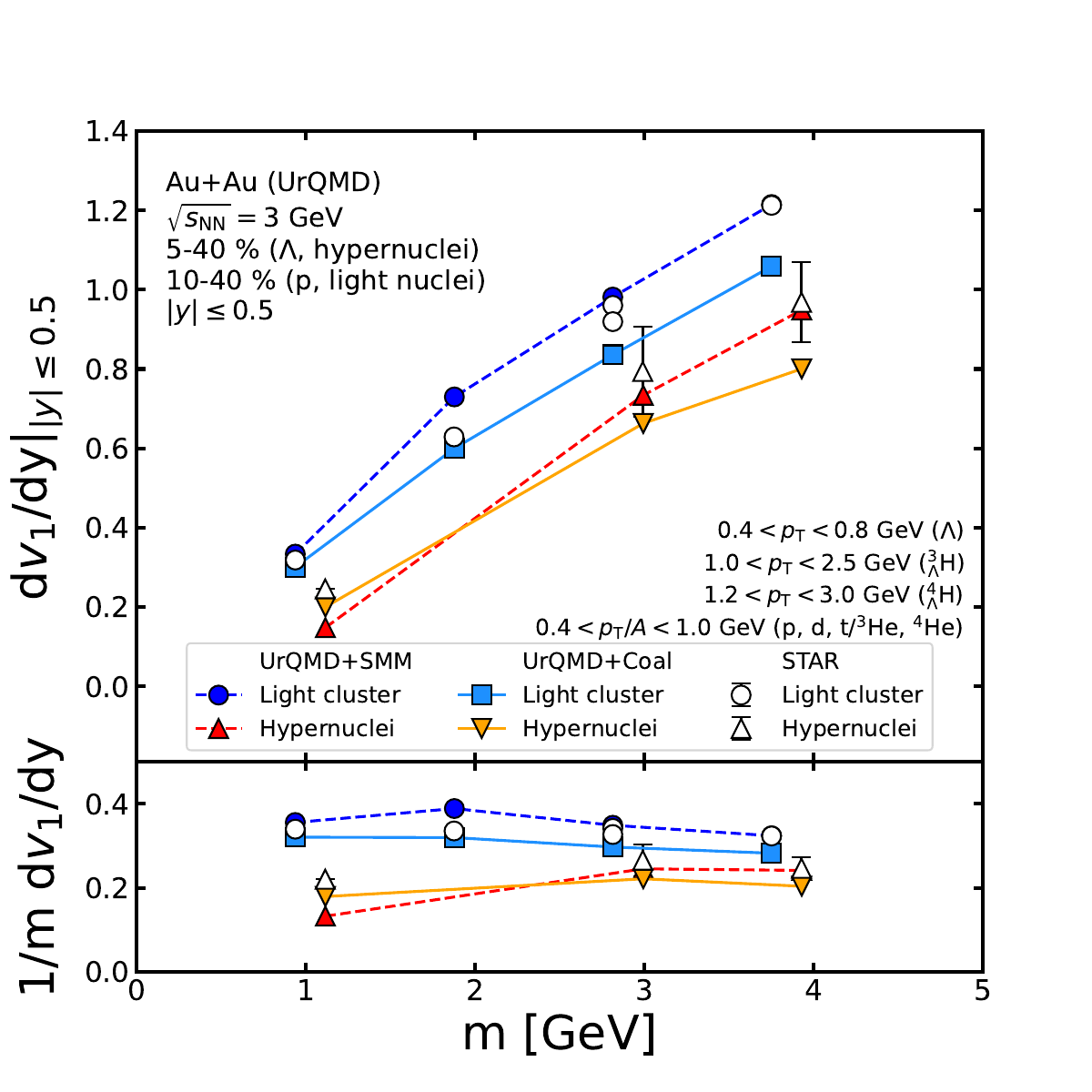}
    \caption{[Color online] The slope of the directed flow $\mathrm{d}v_1/\mathrm{d}y|_{y=0}$ at midrapidity as a function of the mass of the investigated particles from UrQMD with coalescence (light cluster: light blue squares, hypernuclei: yellow triangles-down) and from statistical multi-fragmentation (light cluster: dark blue circles, hypernuclei: red triangles-up) from 5-40\% central (hypernuclei) and 10-40\% central (light nuclei) Au+Au collisions at $\sqrt{s_\mathrm{NN}}=3$. Experimental data points are taken from STAR \cite{STAR:2022fnj}. The lower panel shows the slope of the directed flow scaled by the mass of the particle $1/m \, \mathrm{d}v_1/\mathrm{d}y$ at midrapidity.}
    \label{fig:dv1dy}
\end{figure}

\subsection{Directed and elliptic flow of protons and light nuclei}
We first present the results for the directed and elliptic flow of protons and light nuclei and compare to the experimentally measured data by STAR from Refs. \cite{STAR:2021yiu,STAR:2021ozh,STAR:2022fnj} and then move on to the flow of the hypernuclei in the next section. It should be noted that the transverse momentum ranges used in the STAR publications \cite{STAR:2021ozh}, \cite{STAR:2021yiu} and \cite{STAR:2022fnj} differ slightly. For the following results, we chose  the transverse momentum ranges used in \cite{STAR:2021ozh} for the extraction of the rapidity dependence of $v_1$ and $v_2$ which is $0.4 \leq p_\mathrm{T}/A \leq 1.0$ GeV with $A$ being the mass number of the proton or cluster.

Fig. \ref{fig:v1_y_light} shows the directed flow $v_1$ as a function of rapidity of protons (upper left), deuterons (upper right), tritons and $^3$He (lower left) and $^4$He (lower right) from 10-40\% central Au+Au collisions at $\sqrt{s_\mathrm{NN}}=3$ GeV from UrQMD with coalescence (solid black lines) and from UrQMD combined with the statistical multi-fragmentation model (dashed lines).

The experimental data around midrapidity are well described by both, the coalescence approach and the statistical multi-fragmentation calculation. The SMM calculation shows slightly higher values of the directed flow at large forward and backward rapidities which is favored by the experimental data for the more massive ($A=3,4$) clusters. The protons and deuterons on the other hand seem to favor the coalescence calculation, suggesting an earlier formation of the heavier clusters at mid-rapdity, in line with \cite{Kireyeu:2022qmv}. At larger rapidities, the heavier clusters appear to be formed later or by a different mechanism as in the coalescence which is expected for the fragmentation region. The observed agreement between the data and both model calculations supports the general capability of the employed cluster production models to capture the space-time and space-momentum correlations created in low energy heavy-ion collisions. 

Fig. \ref{fig:v2_y_light} shows the elliptic flow $v_2$ as a function of rapidity of protons (upper left), deuterons (upper right), tritons and $^3$He (lower left) and $^4$He (lower right) from 10-40\% central Au+Au collisions at $\sqrt{s_\mathrm{NN}}=3$ GeV from UrQMD with coalescence (dashed lines) and from UrQMD combined with the statistical multi-fragmentation model (solid lines). Experimental data points are taken from STAR \cite{STAR:2021ozh,STAR:2021yiu}.

The rapidity dependence of the elliptic flow of the protons and the light clusters exhibits the usual shape known in the ``squeeze-out" energy regime, i.e. it follows a concave shape with a negative value at midrapidity. The central value at midrapidity with slightly negative $v_2$ values indicates that shadowing and squeeze-out are still present at a center-of-mass energy of $\sqrt{s_\mathrm{NN}}=3$ GeV. Moving to forward/backward rapidity the elliptic flow increases as expected. While at midrapidity both approaches yield similar results, the statistical multi-fragmentation calculation again shows a stronger increase of $v_2$ at forward/backward rapidities and is slightly more favored by the data over the coalescence calculation.

However, the data for all nuclei seem to indicate a similar value for $v_2$ at midrapidity which is not in qualitative agreement with the simulations.
Both the coalescence and SMM model show an approximate mass number scaling which would not be observed in the data for $v_2$.

\subsection{Directed and elliptic flow of Lambdas and hypernuclei}
After having benchmarked the coalescence and statistical multi-fragmentation results in the last section, we next discuss the results on the flow of strange hadrons and hyperclusters. For the extraction of the directed and elliptic flow of the hyperons and the hypernuclei different transverse momentum cuts are applied in line with the ones used in the STAR publication in Ref. \cite{STAR:2021ozh}. They are $0.4 \leq p_\mathrm{T} \leq 0.8$ GeV for the $\Lambda$, $1.0 \leq p_\mathrm{T} \leq 2.5$ GeV for the $^3_\Lambda$H and $1.2 \leq p_\mathrm{T} \leq 3.0$ GeV for the $^4_\Lambda$H.

Fig. \ref{fig:v1_v2_y_hyper} shows the directed flow $v_1$ (left column) and the elliptic flow $v_2$ (right column) as a function of rapidity of the $\Lambda$ (upper row), $^3_\Lambda$H (middle row) and $^4_\Lambda$H (lower row) from 5-40\% central Au+Au collisions at $\sqrt{s_\mathrm{NN}}=3$ GeV from UrQMD with coalescence (solid lines) and from UrQMD combined with the statistical multi-fragmentation model (dashed lines). Experimental data points are taken from \cite{STAR:2021yiu,STAR:2022fnj}.

The measured directed flow of the $\Lambda$ is in good agreement with both of the model calculations in the whole rapidity range. This suggests that the resulting $\Lambda N$ correlations and the strangeness production channels are correct in the transport model. Although error bars on the measured directed flow of the hypertriton and hyperhydrogen-4 are still large, the general trend of $v_1$ is well reproduced by the UrQMD model with coalescence or with subsequent statistical multi-fragmentation. We additionally point out that we employ momentum dependent potentials for the hyperons that are consistent with the known hyperon potentials in nuclear matter, yielding a similarly good description of the experimental data on hyperon flow as other models that also included a momentum dependent hyperon specific potential \cite{Nara:2022kbb}.

Turning to the elliptic flow in the right column of Fig. \ref{fig:v1_v2_y_hyper}, the $\Lambda$ also shows a concave rapidity dependence around midrapidity with a slightly negative value at $y\approx0$. The full UrQMD simulation supplemented with the coalescence approach and the statistical multi-fragmentation calculation both match the data qualitatively well and show a slightly increasing elliptic flow at forward and backward rapidities, in line with the data\footnote{We point out that the $v_2$ of the $\Lambda$ is sensitive to the onset of inelasticity of the $\sigma_{p\Lambda}$ cross section.}. The $v_2$ of the $^3_\Lambda$H exhibits a minimum at midrapidity and increases more strongly towards forward and backward rapidity as it has been seen in the light clusters before. Here the statistical multi-fragmentation and the coalescence approach agree with each other. The $^4_\Lambda$H lacks decent statistics, but the trend of the curves seems to coincide broadly with the hypertriton depicting a concave shape as well. Both the hypernuclei rather follow the flow pattern dictated by the deuteron (triton) part of the bound object. The measurement of the elliptic flow of hypernuclei may therefore allow to constrain the source size of hypernuclei shedding light onto the production mechanism of those objects and further constrain the $YN$ interaction more precisely.

The measurement of (higher order) harmonic flow of hypernuclei will benefit from the enormous luminosity of the upcoming FAIR facility allowing for more detailed comparisons and model improvements.

\begin{figure}[t]
    \centering
    \includegraphics[width=\columnwidth]{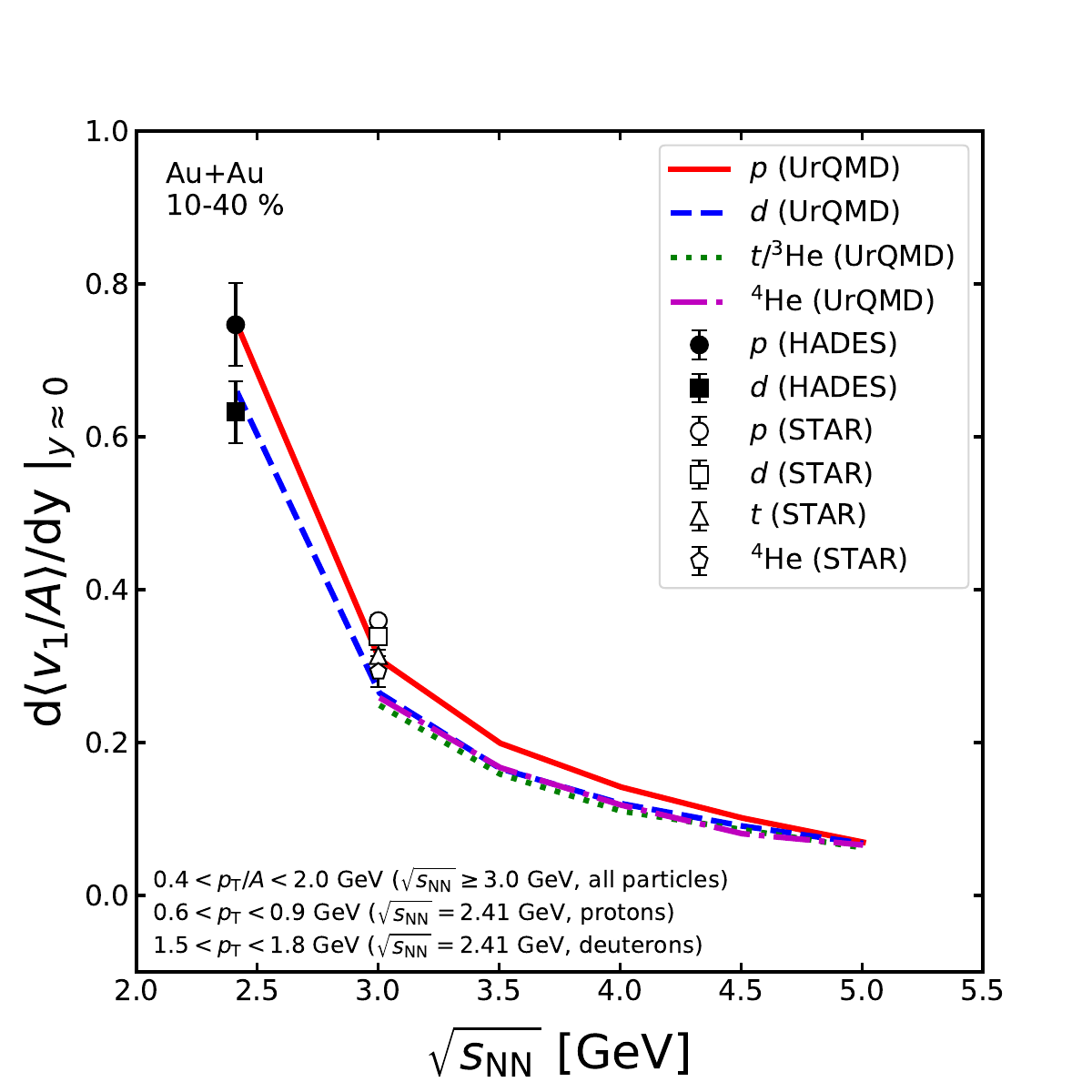}
    \caption{[Color online] The slope of the directed flow $\mathrm{d}(v_1/A)/\mathrm{d}y|_{y=0}$ at midrapidity scaled by the mass number as a function of center-of-mass energy from UrQMD with coalescence in 10-40\% central Au+Au collisions. Experimental data points are taken from \cite{HADES:2022osk,STAR:2021yiu}.}
    \label{fig:dv1dy_sNN}
\end{figure}
\begin{figure}[t]
    \centering
    \includegraphics[width=\columnwidth]{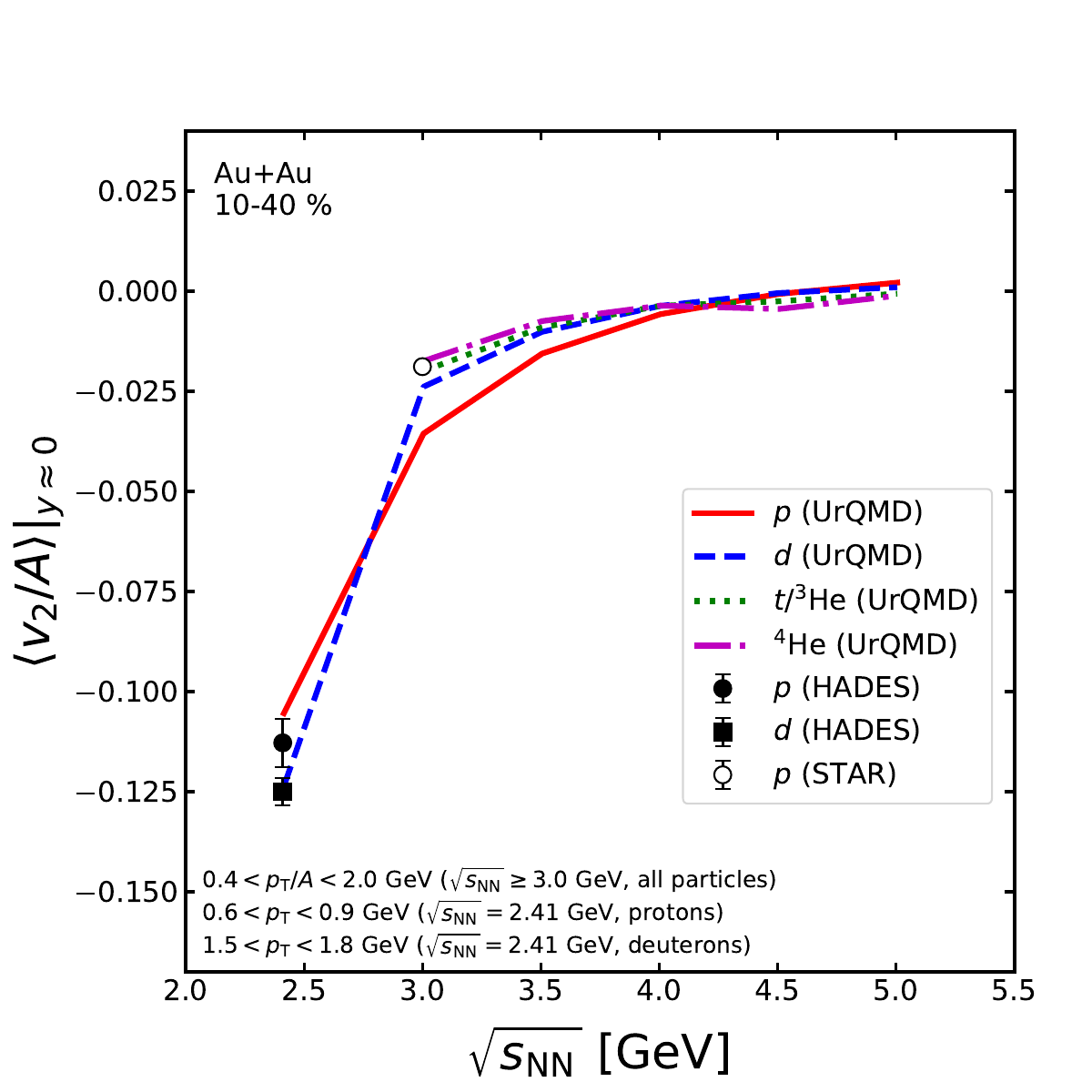}
    \caption{[Color online] The elliptic flow $v_2/A|_{y=0}$ at midrapidity scaled by the mass number as a function of center-of-mass energy from UrQMD with coalescence in 10-40\% central Au+Au collisions. Experimental data points are taken from \cite{HADES:2022osk,STAR:2021yiu}.}
    \label{fig:v2_sNN}
\end{figure}

\subsection{Slope of the directed flow at midrapidity}
The slope of the directed flow at midrapidity is sensitive to the incompressibility of the Equation-of-State. We therefore provide in Fig. \ref{fig:dv1dy} the slope of the directed flow $\mathrm{d}v_1/\mathrm{d}y|_{y=0}$ at midrapidity as a function of the mass of the investigated particles from UrQMD with coalescence (light cluster: light blue squares, hypernuclei: yellow triangles-down) and from statistical multi-fragmentation (light cluster: dark blue circles, hypernuclei: red triangles-up) from 5-40\% central (hypernuclei) and 10-40\% central (light nuclei) Au+Au collisions at $\sqrt{s_\mathrm{NN}}=3$. 
Experimental data points are taken from STAR \cite{STAR:2022fnj}.

The slope of the directed flow of the protons and $\Lambda$'s is essentially equal in both the coalescence and statistical multi-fragmentation calculation. For larger clusters, however, the results from the statistical multi-fragmentation calculations are systematically larger than the coalescence results and agree better with the experimental data for larger nuclei. The reason is, as discussed before, the better description of the fragmentation region which is more important for the larger clusters. In case of the hypernuclei, both the coalescence and statistical multi-fragmentation calculation match the experimental data very well. One generally observes an almost linear scaling behavior of $\mathrm{d}v_1/\mathrm{d}y|_{y=0}$ with the mass $m$ or the mass number $A$ respectively indicating that they are governed by the same underlying velocity field. Note, that small deviations from the exact scaling behavior are expected due to the fact that the light clusters constitute a significant fraction of the total baryon number and therefore the measurable proton and hyperon flow is not equal exactly to the flow of protons and hyperons before the cluster formation.

\subsection{Predictions for mass number scaling}
Figures \ref{fig:dv1dy_sNN} and \ref{fig:v2_sNN} show our coalescence predictions for the scaling of directed and elliptic flow of protons, deuterons, triton and $^4$He, for the whole energy range covered by the STAR fixed target program and the upcoming CBM experiment at FAIR. In both cases, the slope of the directed flow and the integrated elliptic flow are scaled by the mass number $A$ of the light nuclei. For comparison we also included a comparison with available HADES data \cite{HADES:2022osk}. While the HADES data are almost perfectly described by the simulations with coalescence we observe slight deviations from the STAR data as discussed above. 
More importantly, while at $\sqrt{s_{\mathrm{NN}}}\le 3$ GeV the mass scaling is not perfect, the scaling improves significantly for higher beam energies. This will allow for a much better extraction of the formation properties of light nuclei in the future.

\section{Conclusion}
The Ultra-relativistic Quantum Molecular Dynamics (UrQMD) model supplemented with a coalescence mechanism and the statistical multi-fragmentation (SMM) model has been employed to calculate light cluster and hypernuclei flow in 5-40\% (and 10-40\%) peripheral Au+Au collisions at $\sqrt{s_\mathrm{NN}}=3$ GeV. The calculated directed flow $v_1$ of the protons, $\Lambda$s, the light cluster and the hypernuclei show a quantitative agreement with the recent measurements by STAR. The mass number scaling is approximately observed with small deviations that are sensitive to the formation time of the clusters closer to the fragmentation regions.

The elliptic flow $v_2$ of the protons, $\Lambda$ agrees with the experimental data. In contrast to the STAR experiment we observe an approximate mass scaling for $v_2$ at midrapidity. 

Based on both production models, the elliptic flow of the $^3_\Lambda$H and $^4_\Lambda$H for $\sqrt{s_{\mathrm{NN}}}\le 3$ GeV has been predicted. Their precise measurement by the upcoming CBM experiment at FAIR will allow to understand the hyperon-nucleon interaction more precisely.

In addition, we have predicted the scaling of the directed and elliptic flow for the whole energy range of the STAR fixed target program and the upcoming CBM experiment. 

\begin{acknowledgements}
The authors thank Benjamin D\"onigus, Manuel Lorenz, Behruz Kardan and Christoph Blume for fruitful discussion about the flow harmonics and hypernuclei. 
We further thank Elena Bratkovskaya and J\"org Aichelin for inspiring discussions about clustering and the generation of harmonic flow.
T.R. acknowledges support through the Main-Campus-Doctus fellowship provided by the Stiftung Polytechnische Gesellschaft (SPTG) Frankfurt am Main.
T.R. and J.S. thank the Samson AG for their support. 
M.O.K. was supported by the ErUM-Data funding line from BMBF through the KISS project. 
This project was supported by the DAAD (PPP T\"urkiye). 
N.B. thanks the Scientific and Technological Research Council of T\"urkiye (TUBITAK) for their support under Project No. 124N105.
Computational resources were provided by the Center for Scientific Computing (CSC) of the Goethe University and the ``Green Cube" at GSI, Darmstadt. 
\end{acknowledgements}



\end{document}